\documentclass[twocolumn,secnumarabic,amssymb, nobibnotes, aps,nofootinbib, prd]{revtex4-1}
\usepackage{amsmath}
\usepackage{myterms}
\usepackage{color}
\usepackage{graphicx}
\usepackage{multirow}

\def\Mpl{M_{\rm pl}}
\def\lsim{\mathrel{\rlap{\lower4pt\hbox{\hskip1pt$\sim$}}
     \raise1pt\hbox{$<$}}}
\def\gsim{\mathrel{\rlap{\lower4pt\hbox{\hskip1pt$\sim$}}
     \raise1pt\hbox{$>$}}}

\newcommand{\mc}{\mathcal}
\newcommand{\bv}[1]{{\bf #1}}

\begin{document}

\title{
B-modes from Post-inflationary  Gravitational Waves Sourced by Axionic Instabilities at Cosmic Reionization
}
\author{Michael Geller}
\email{mic.geller@gmail.com}
\author{Sida Lu}
\email{sidalu@mail.tau.ac.il}
\affiliation{Department of Physics, Tel Aviv University, Tel-Aviv 69978, Israel}

\author{Yuhsin Tsai}
\email{ytsai3@nd.edu}
\affiliation{Department of Physics, University of Notre Dame, South Bend, IN 46556, USA}	

\begin{abstract}
We show that axion-like particles  that only couple to invisible dark photons can generate visible B-mode signals around the reionization epoch. The axion field starts rolling shortly before reionization, resulting in a tachyonic instability for the dark photons. This generates an exponential growth of the dark photon quanta sourcing both scalar metric modes and gravitational waves that leave an imprint on the reionized baryons. The tensor modes modify the cosmic microwave background (CMB) polarization at reionization, generating visible B-mode signatures for the next generation of CMB experiments for parameter ranges that satisfy the current experimental constraints.
\end{abstract}

\maketitle

\section{Introduction}\label{sec:introduction}
The discovery of gravitational waves (GW) at LIGO~\cite{aasi2015advanced} and VIRGO~\cite{acernese2014advanced} has motivated the search for other possible GW sources beyond the mergers of astrophysical objects. Among those, GWs from cosmological sources, such as strong first order phase transitions~\cite{Caprini:2019egz} and the presence of cosmic strings~\cite{Auclair:2019wcv}, are of particular interest in elucidating the early history of the universe (e.g.,~\cite{Geller:2018mwu,Cui:2017ufi}). The cosmological GW signals can have a wide range of possible frequencies: interferometer experiments can detect GWs with frequencies above $\sim10^{-5}~$Hz~\cite{Kawamura:2011zz,2017arXiv170200786A,Punturo:2010zz,Crowder:2005nr}, and lower frequency signals down to $\sim 10^{-8}~$Hz are relevant for pulsar timing experiments~\cite{dewdney2009square,manchester2013international}; if GWs have frequencies lower than $\sim 10^{-15}~$Hz, we can search for the B-mode polarization signals from GW imprints on the cosmic microwave background (CMB)~\cite{Ade:2015fwj}. Such low-frequency signals have wavelengths comparable to the visible universe's size and must have a cosmological origin. As a result, the B-mode signal is mainly considered to come from GWs produced during cosmic inflation (see~\cite{Kamionkowski:2015yta} and the references therein). 

In this letter, we propose a new source for B-mode generating GWs produced by axion-like particles (ALPs) around the time of cosmic reionization. Axions were originally proposed to solve the strong CP problem~\cite{Peccei:1977hh,Peccei:1977ur} and realized to be a viable dark matter (DM) candidate~\cite{Abbott:1982af,Preskill:1982cy,Dine:1982ah,Co:2017mop}. ALPs generalize the cosmological phenomenology of axions without a necessary connection to strong CP. For example, an ALP can serve as the inflaton field responsible for the period of cosmic inflation~\cite{Freese:1990rb,Dimopoulos:2005ac,Anber:2009ua} or as the relaxion, addressing the hierarchy problems in nature by varying the fundamental constants of nature with time ~\cite{Hook:2016mqo,Fonseca:2019ypl,Graham:2015cka}. On the experimental side, several new direct detection experiments have been put into action~\cite{Anastassopoulos:2017ftl,Du:2018uak,Ouellet:2018beu,Zhong:2018rsr} or have been proposed~\cite{Liu:2018icu,Bogorad:2019pbu,Hook:2018iia,Caputo:2018vmy} to look for ALPs. Part of the theoretically-favored axion parameter space has already been experimentally excluded.

In the particular case where ALPs couple to dark photons, the ALP field's rolling leads to a ``tachyonic instability" that amplifies vacuum fluctuations of one of the dark photon helicities. The process generates exponential dark photon production, and similar phenomena have been studied under the context of inflation~\cite{Anber:2009ua,Anber:2012du}, production of dark photon DM~\cite{Co:2018lka}, depletion of axion DM to avoid overclosure~\cite{Agrawal:2017eqm}, and friction for the relaxion models~\cite{Hook:2016mqo,Fonseca:2019ypl}. Recently it has been shown that the stochastic GW background generated through this process in the early universe may be detectable in interferometers or pulsar timing arrays~\cite{Machado:2018nqk,Machado:2019xuc}. In~\cite{Weiner:2020sxn}, a similar mechanism at the recombination period is studied within the context of early dark energy solutions to the Hubble tension~\cite{Bernal:2016gxb,Poulin:2018cxd} and is shown to produce visible GW signals in the CMB.

In this work, we consider a similar effect of producing a GW background from tachyonic particle production late in the universe's history -- after recombination and around the time of galaxy formation.
As a tensor perturbation of the metric, the GW background leaves an imprint on the photon energy distribution. When the universe enters the reionization era at $z_{\rm rei}\approx 8$, CMB photons propagating in the line-of-sight direction get polarized by the last Thomson scattering, and a combination of the tensor  perturbation and the angular distribution of the photon polarization produces the B-mode signal in the large-scale CMB spectrum. In particular, we will show that for parameter ranges of our model not currently excluded by existing or past experiments \cite{Aghanim:2019ame,Ade:2015tva}, we predict a B-mode signal accesible to the next generation of B-mode detectors. 

The B-mode signals sourced by the axionic instability have a power spectrum which could be distinguished from those produced by inflationary GWs. An observation of such unique B-mode signals will be a discovery of dark sector physics and will shed light on the nature of dark energy by revealing that dark energy is changing at late times. In particular, a revelation that dark energy has recently changed by an amount close to its current value is suggestive of some dynamics related to the cosmological constant (CC) problem. As we will show, next-generation experiments will be able to probe such shifts on a scale similar to the current value of the cosmological constant. 

We remark that our calculation utilizes a linear semi-classical approximation and therefore our results need to be confirmed by a full lattice study. We expect this to affect the precise predictions of the spectral shapes, but not our ultimate conclusions.

This paper is organised as follows: After reviewing the mechanism of tachyonic production of dark photons, we describe our setting and set up the calculation of dark photon's energy density fluctuations in Sec.~\ref{sec:model}. We then discuss the metric perturbation sourced by the dark photon fluctuation in Sec.~\ref{sec:CMB} and show the derivation of the resulting CMB spectra.
Subsequently, we present our results, comparing the predicted signals within two benchmark ALP models to the sensitivity of the future B-mode experiments and to the current constraints from Planck in Sec.~\ref{sec:CMB_results}. Finally, we conclude in Sec.~\ref{sec:Conclusion}. 

\section{The Model}\label{sec:model}

\subsection{Tachyonic production of dark photons}\label{sec:tachyonic}

We consider an axion field $\phi$ coupled to a $U(1)$ dark photon, with
the Lagrangian given by
\begin{align}
\mc{L}=-\dfrac{1}{2}\partial_\mu\phi \partial^\mu\phi-V(\phi)-\dfrac{1}{4}F_{\mu\nu}F^{\mu\nu}-\dfrac{\alpha}{4f}\phi F_{\mu\nu}\tilde{F}^{\mu\nu}\,,
\end{align}
where $V(\phi)=\frac{1}{2}m^2\phi^2$, and $f$ is the axion constant. We assume the dark photon is massless and is produced only after inflation. The quadratic potential $V(\phi)$ can naturally arise from an axion-like potential $\Lambda^4\cos(\phi/f)$, which implies $m\sim \Lambda^2/f$. We consider $m$ close to the Hubble scale right before the reionization.
We will see that in our setting, the CMB probes $\Lambda\sim {\cal O}({\rm meV})$, which also coincides with the order of magnitude of the cosmological constant, so that an observation of the signal we discuss may lead to new insights into dark energy \footnote{For example, ~\cite{Graham:2019bfu} has proposed a similar axion model to address the cosmological constant problem.}.

The equation of motion of the axion field is then
\begin{align}\label{eq:axion_eom}
\phi^{\prime\prime}+2aH\phi^\prime+a^2\frac{\partial V}{\partial \phi}=\frac{\alpha}{f}a^2\bv{E}\cdot\bv{B}\,,
\end{align}
in which $a$ is the scale factor of the FRW metric $ds^2=a^2(\tau)(d\tau^2-\delta_{ij}dx^idx^j)$, and $H$ is the Hubble parameter. 
The prime symbol denotes derivatives with respect to the conformal time $\tau$. 
On the right hand side of the equation, the dark electromagentic field serves as friction for the rolling of the axion $\phi$.

The rolling of the axion will cause the dark photon modes within a certain momentum range to grow exponentially, a phenomenon known as the tachyonic instability. 
This can be shown by examining the equation of motion of the dark photon field, which in the Coulomb gauge is written as
\begin{equation}\label{eq:dark_photon}
\begin{aligned}
&X_i = \int \mathcal{D}k\left(\epsilon_{+i}(\bv{k})v_+(\tau,k)\hat{{\bf a}}_+(\bv{k})e^{i\bv{k}\cdot\bv{x}}+h.c.\right),\\
&X_0 = 0\,,
\end{aligned}
\end{equation}
where $\mathcal{D}k\equiv d^3k/(2\pi)^3$. The creation and annihilation operators obey the commutation relation $[{\bf a}_+(\bv{k}),\,{\bf a}^\dagger_+(\bv{k^\prime})] = (2\pi)^3\delta(\bv{k}-\bv{k}^\prime)$, and the polarization vectors obey $\bv{k}\cdot\bv{\epsilon}_\pm=0$, $\bv{k}\times\bv{\epsilon}_\pm=\mp i k\bv{\epsilon}_\pm$, $\bv{\epsilon}_\pm\cdot\bv{\epsilon}_\pm=0$, $\bv{\epsilon}_\pm\cdot \bv{\epsilon}_\mp=1$. 
The dark photon field equation can then be written in terms of the mode function $v$ as
\begin{align}
v^{\prime\prime}_\pm(k,\tau)+\omega^2_\pm(k,\tau) v_\pm(k,\tau)=0,
\end{align}
with the dispersion relation $\omega^2_\pm(k,\tau)=k^2\mp k\alpha\phi^\prime/f$. As long as the axion starts rolling and develops a non-zero $\phi^\prime$, the dark photon modes of a certain helicity in the momentum band $0<k<\alpha\lvert\phi^\prime\rvert/f$ will have $\omega^2(k,\tau)<0$ and therefore grow exponentially. 
Specifically, the $v_+$ modes can grow when $\phi^\prime>0$ and the $v_-$ modes grow when $\phi^\prime<0$, and the growth of the two helicities are alternating as the axion field oscillates around the minimum of its potential. 
The helicity experiencing the tachyonic instability right after axion starts the rolling will be significantly more enhanced than the other, as it spends more time in the tachyonic band. 

To solve the axion and the dark photon coupled equations of motion, we treat the dark photon mode functions $v_\pm(k,\tau)$ as discretized modes of fixed $k$. 
And to the leading order, the reaction from dark photon field $\bv{E}\cdot\bv{B}$ on the right hand side of \eref{eq:axion_eom} is replaced by the expectation value $\langle\bv{E}\cdot\bv{B}\rangle$, which is calculated as
\begin{equation}\label{eq:EdotB}
\langle\bv{E}\cdot\bv{B}\rangle =-\sum_{\lambda=\pm}\lambda \int \dfrac{k^2dk}{2\pi^2 a^4}\,{\rm Re}\left[v^\ast_\lambda(k,\tau)v^\prime_\lambda(k,\tau)\right]\,.
\end{equation}

\subsection{Calculation setup}\label{sec:setup}
\begin{figure*}[thb]
\centering
\includegraphics[height=0.31\linewidth]{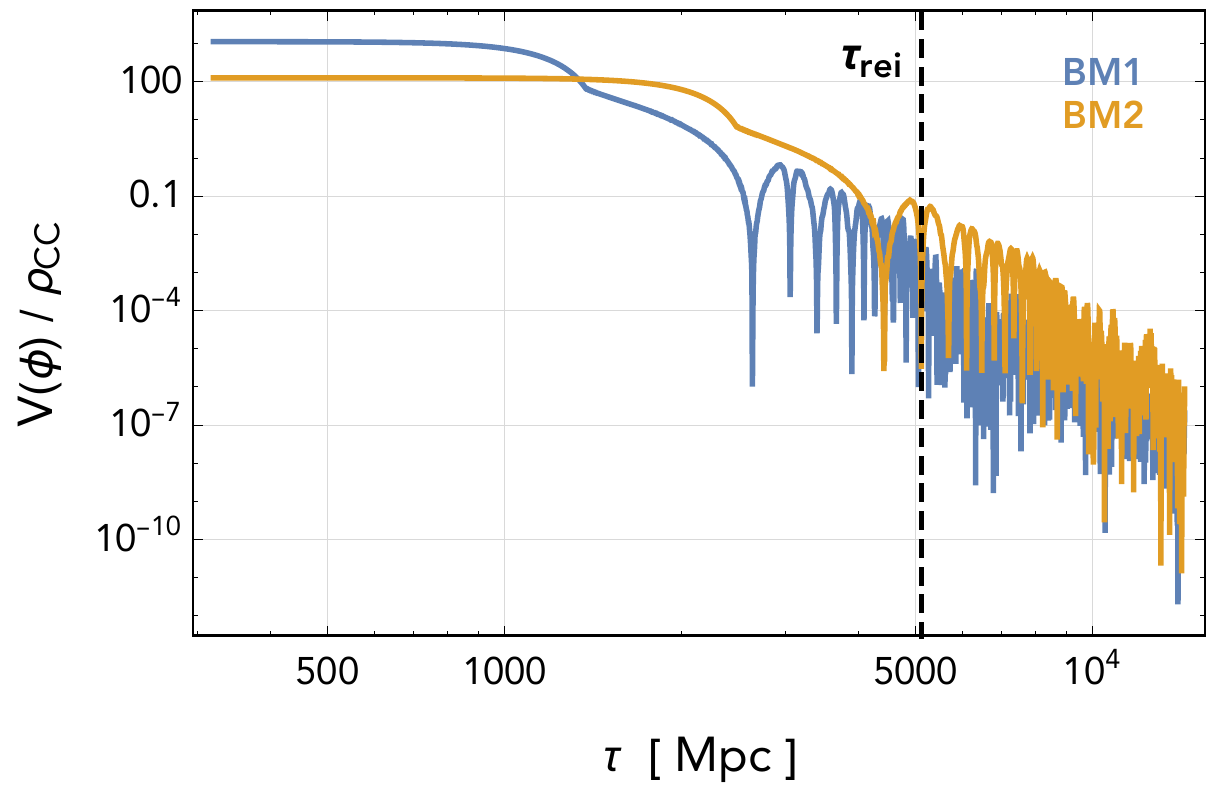}\qquad 
\includegraphics[height=0.31\linewidth]{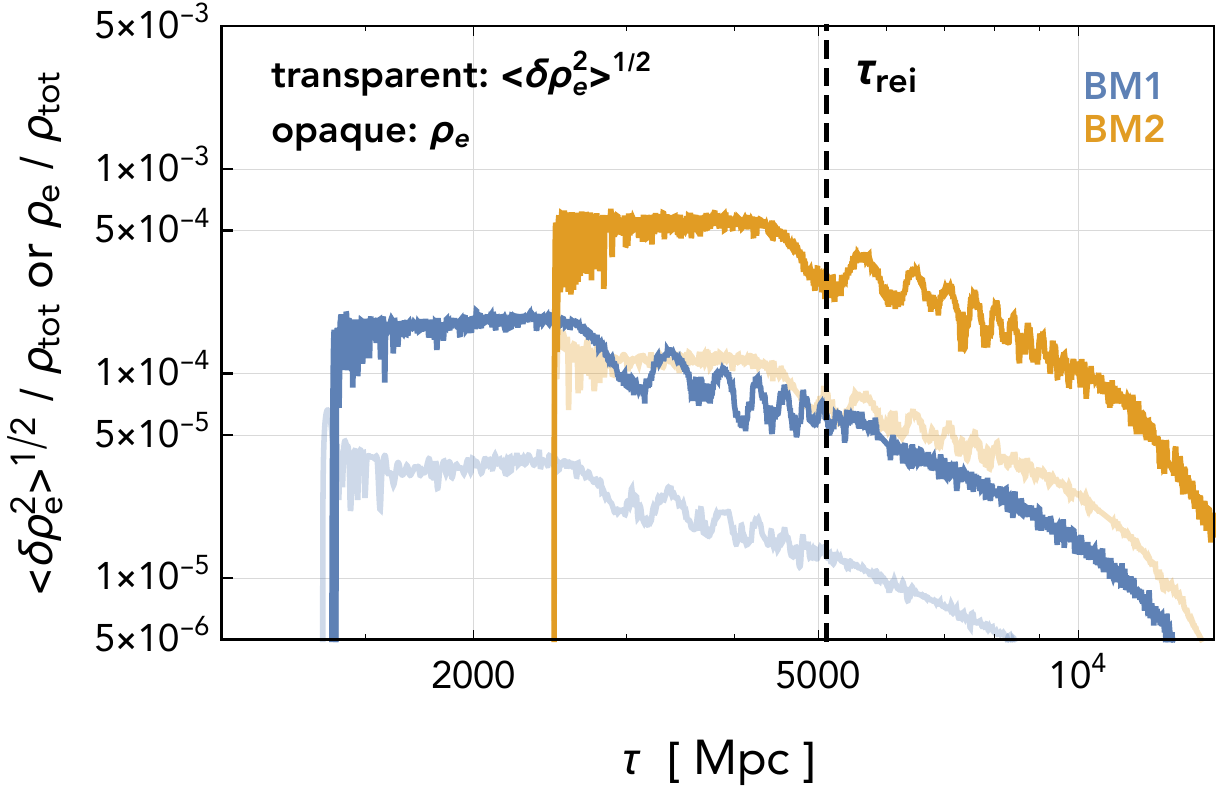}
\caption{
{\it Left panel}: The evolution of the axion potential energy of the two benchmark models we use, normalized by the dark energy density today $\rho_{CC}\approx 37~{\rm meV}^4$.
{\it Right panel}: The evolution of the dark sector energy density $\rho_e$ and its perturbation $\langle\delta\rho^2_e\rangle^{1/2}$ induced by the tachyonic particle production, normalized by the total energy density of the universe, where $\langle\delta\rho^2_e\rangle^{1/2}\equiv\left[\int\mc{D}k\mc{D}k^\prime\langle\delta\rho_e(k)\delta\rho_e(k^\prime)\rangle\right]^{1/2}$.
}
\label{fig:V_over_Lambda}
\end{figure*}

In our calculation, we assume the dark photon to be non-thermal such that its abundance comes only from the tachyonic production described above. 
We use 200 dark photon k-modes equally spaced on a logarithmic grid in the  momentum range $[k_{\rm min}, k_{\rm max}]$. 
The value of $k_{\rm max}$ is chosen such that $k_{\rm max}\gtrsim \alpha \lvert \phi^\prime\rvert_{\rm max}/f$, and we perform a consistency check with several choices of $k_{\rm max}$ to determine the number used for each calculation. The minimum value of the momentum range is set to be $k_{\rm min}=H_0/4$. With these choices, we make sure that the entirety of the momentum range of interest is covered. 

In this work we do not include the back reaction of the gauge modes on the axion perturbations that requires a full lattice study (see \cite{Ratzinger:2020oct}). This is mainly important for the axion abundance calculation which is not of interest in this setup. For the GWs, we expect the lattice results to be roughly consistent in magnitude \cite{Ratzinger:2020oct} and to be mainly important for the spectral shape (see also \cite{Kitajima:2017peg,Agrawal:2018vin,Kitajima:2020rpm}). We therefore treat the calculation here as a preliminary estimate to motivate a full lattice study which is left for a subsequent work.

The produced dark photon k-modes are assumed to be in the Bunch-Davis vacuum $v_\pm(k,\tau)=e^{ik\tau}/\sqrt{2k}$ before the axion rolling starts and
the axion field is released at an initial misalignment of $\lvert \phi_0 \rvert=f$. We choose two benchmark models for which the tachyonic instability becomes significant after recombination but before reionization, taken as $z_{\rm rei}=8$. Note that keeping the axion mass fixed and varying the height of the initial misalignment ({\it i.e.} $\Lambda$ in our parameterization, keeping $f\propto \Lambda^2$) will only rescale the energy in the dark sector, and with it the energy density in the perturbations. Therefore we will think of the two benchmarks as two classes of models where the total energy in the dark sector remains a free parameter which can be constrained by current experimental data.
We give the benchmark values of the parameters in Table~\ref{tab:para}, where we also show bounds on the energy scale in the dark sector $\Lambda_{\rm bound}$ which we derive in the following sections based on the uncertainty of the current power spectra measurements.

We plot the time evolution of the axion potential and the resulting inhomogeneities in the gauge modes in Fig.~\ref{fig:V_over_Lambda}, using the two benchmark mark models and energy scale $\Lambda_{\rm bound}$. We see that the axions start their rolling shortly before the reionization and begin to oscillate around the minimum, producing the gauge mode inhomogeneity in the process. In Fig.~\ref{fig:V_over_Lambda} (right), we see that the dark photon energy is always below ${\cal O}(5\times 10^{-4})$ of the total energy. The process therefore gives negligible corrections to the angular diameter distance that relates to the CMB spectra. However, even though the average $\rho_e$ is small comparing to $\rho_{\rm tot}$ that is dominated by the matter density $\rho_m$, the density contrast of the dark photon energy is of $\mathcal{O}(1)$ as can be seen in the transparent and opaque curves. The energy perturbation $\langle\delta\rho_e^2\rangle^{1/2}$ is thus comparable to the matter density perturbation ($\sim 10^{-5}\rho_m$) that enters the horizon around the same time and can therefore generate visible signals as we show below.

\begin{table}[htb]
\centering
\renewcommand{\arraystretch}{1.2}
\setlength\tabcolsep{0.8em}
\begin{tabular}{|c|c|c|c|c|c}
\hline\hline
	& $m$ (eV)&  $k_{\rm max}$ (Mpc$^{-1}$)& $\Lambda_{\rm bound}$ & $\alpha$\\
\hline
BM1 & $4\times 10^{-30}$ & 0.94 & 16 meV & 400\\
BM2 & $8.8\times 10^{-31}$ & 0.78 & 9 meV& 400\\
\hline
\hline  
\end{tabular}
\caption{The benchmark parameters used in the calculation. 
}
\label{tab:para}
\end{table}

\section{CMB spectra calculation}\label{sec:CMB}

Although the axion starts rolling only after recombination,  remarkably it can still modify the CMB perturbation observed today. The dark photon field enhanced by the tachyonic instability generates isocurvature perturbations that also source GWs~\cite{Machado:2018nqk} affecting the CMB power spectrum through the late integrated Sachs-Wolfe (ISW) effect. The produced GWs also leave an imprint in the CMB B-mode which will serve as our target signal for the discovery of this setup. Here we present the calculation of CMB $TT$, $EE$ and $BB$ spectra, $C_{\ell}^{TT}$, $C_{\ell}^{EE}$ and $C_{\ell}^{BB}$. 

\subsection{Scalar mode contribution }\label{sec:scalar_pert}
Perturbations of the axion and dark photon energy density $\delta\rho_e$ generate a gravitational potential $\Phi$ through the linear Boltzmann and Einstein equations~\cite{Ma:1995ey}
\begin{equation}\label{eq:linear_ein}
\begin{aligned}
& \delta^\prime_m+\theta_m=3\Phi^\prime\,,\\
& \theta^\prime_m+\dfrac{a^\prime}{a}\theta_m=-\Phi\,,\\
& k^2\Phi+3\dfrac{a^\prime}{a}\Phi^\prime+3\left(\dfrac{a^\prime}{a}\right)^2\Phi=-4\pi G_N\,a^2(\delta\rho_e+\delta\rho_m)\,.
\end{aligned}
\end{equation}
Here $\delta_m=\frac{\delta\rho_m}{\rho_m}$ is the matter energy density contrast induced by the perturbation from the dark photons, and $\theta_m$ is the velocity divergence of matter. For the metric perturbations we set $\Phi=-\Psi$ and ignore the stress tensor from the free streaming radiation. Once the tachyonic production starts, the dark photons dominate the energy perturbation of the dark sector, and hence:
\begin{align}\label{eq:delta_rho}
\delta\rho_e\approx\frac{1}{2}\frac{1}{a^4(\tau)}\delta\left[\left(\partial_0X_i\right)^2\right]+\frac{1}{4}\delta\left[X^{ij}X_{ij}\right].
\end{align}
Here the energy density fluctuation is defined as an operator by subtracting the expectation value from the energy density operator~\cite{Anber:2009ua}.  

Through the ISW effect, the gravity perturbation $\Phi$, obtained by solving Eq.~(\ref{eq:linear_ein}), sources a temperature perturbation today $\Theta_0(\bv{n})=\delta T/T(-\bv{n};\tau_0)$ as~\cite{gorbunov2011introduction}
\begin{align}
\Theta_0(\bv{n})&=\sum_l i^l\,(2l+1)\int \mc{D}k\,\tilde{\Theta}_l(\bv{k})P_l\left(\dfrac{\bv{k}\cdot\bv{n}}{k}\right),\\
\label{eq:Theta_tilde}
\tilde{\Theta}_l(\bv{k})&=2\int^{\tau_0}_{\tau_{rec}}d\tau\,\Phi^\prime(\bv{k},\tau)j_l[k(\tau_0-\tau)]\,,
\end{align}
where $\tau_0$ and $\tau_{rec}$ are the conformal time today and at recombination respectively. 
Since the dark photon perturbation from the tachyonic production is uncorrelated with the adiabatic perturbation, the cross correlator between $\Theta_0(\bv{n})$ and the adiabatic CMB temperature perturbation is negligible. Therefore, the dark photon contribution to the temperature power spectrum is calculated as
\begin{align}
C^{TT}_l=\dfrac{1}{4\pi}\int d\bv{n^\prime}d\bv{n^{\prime\prime}}\Theta_0(\bv{n^\prime})\Theta_0(\bv{n^{\prime\prime}})P_l(\bv{n^{\prime}}\cdot\bv{n^{\prime\prime}})\,.
\end{align}
Using functions $T^r_l$ and $T^i_l$ defined in \eref{eq:Tr} and \eqref{eq:Ti} as convolution integrals between the dark photon mode function $v(k,\tau)$ and the spherical Bessel functions, we find that
\begin{eqnarray}\label{eq:ClTT_final}
C^{TT}_l=8\pi^3 G^2_N &\displaystyle{\int}\mc{D}k\int\mc{D}k_1 (k_1^2+2k_1k_2+k_2^2-k^2)^2\nonumber \\
&\cdot \left[T^{r2}_{\ell}(k,k_1,k_2)+T^{i2}_{\ell}(k,k_1,k_2)\right]\,,\,
\end{eqnarray}
where the vector $\bv{k_2}=\bv{k}-\bv{k_1}$. We give more details of the derivation in the Appendix~\ref{app:CTT}. 

The scalar perturbations can also source the CMB E-mode, which can be calculated as
\begin{align}
C^{EE}_l =& \dfrac{9\pi}{2}\mc{T}^2_{\rm rei}\dfrac{(l+2)!}{(l-2)!}\int \mc{D}k\mc{D}k^\prime \langle\Phi(\tau_{\rm rei})\Phi(\tau_{\rm rei})\rangle\nonumber\\
&\cdot j^2_2(k\tau_{\rm rei})\cdot\dfrac{j^2_l[(\tau_0-\tau_{\rm rei})k]}{k^4(\tau_0-\tau_{\rm rei})^4}\,
\end{align}
after taking the narrow width approximation of the visibility function in time. Here $\tau_{\rm rei}$ is the conformal time at reionization, and $\mathcal{T}_{\rm rei}\approx 0.08$ is the photon optical depth in the reionized universe. 
We find that the E-mode contribution from the scalar perturbations is subdominant to that of the tensor perturbations.

\subsection{Tensor mode contribution}\label{sec:b-mode}

\begin{figure*}[thb]
\centering
\includegraphics[width=0.46\linewidth]{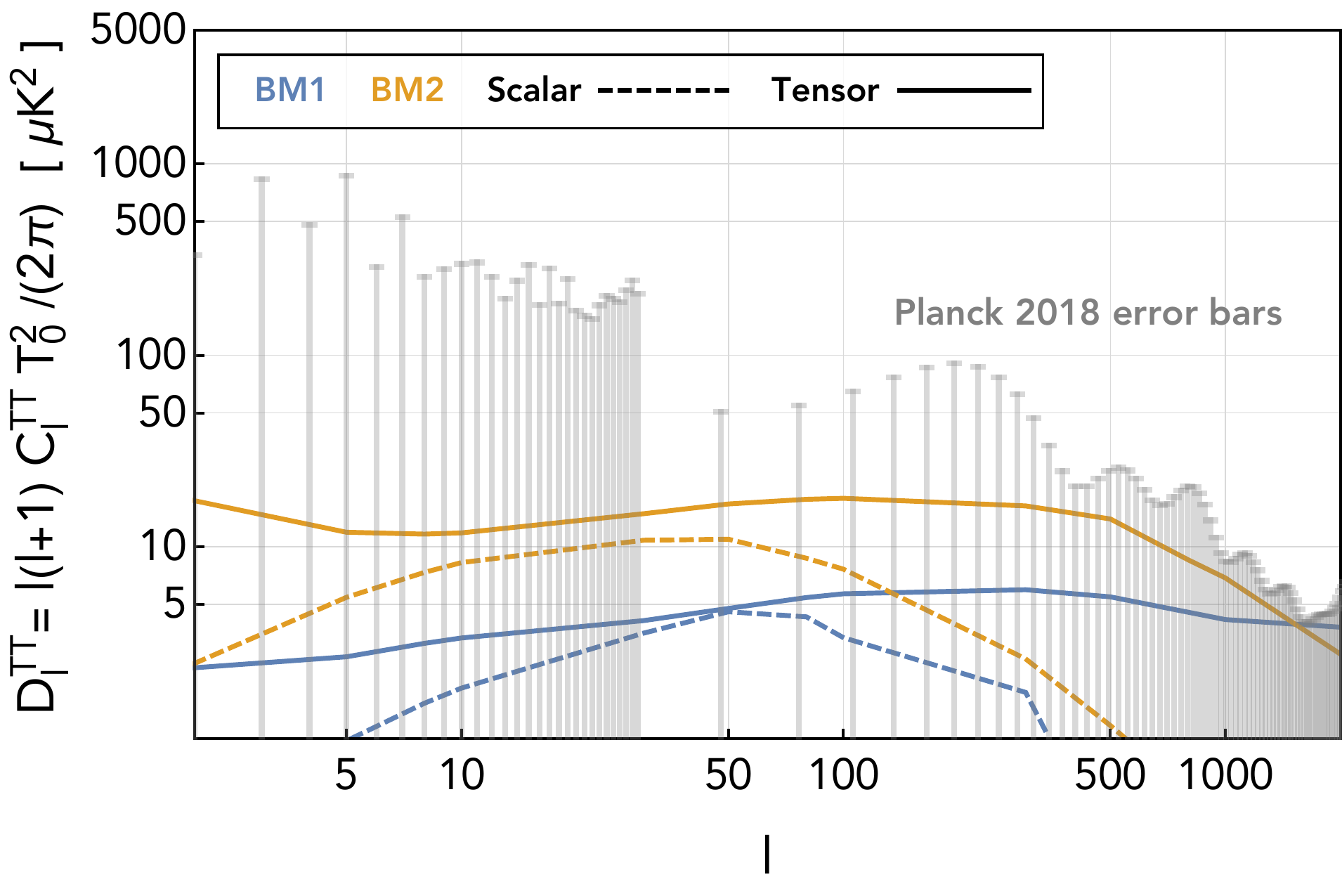} \qquad 
\includegraphics[width=0.468\linewidth]{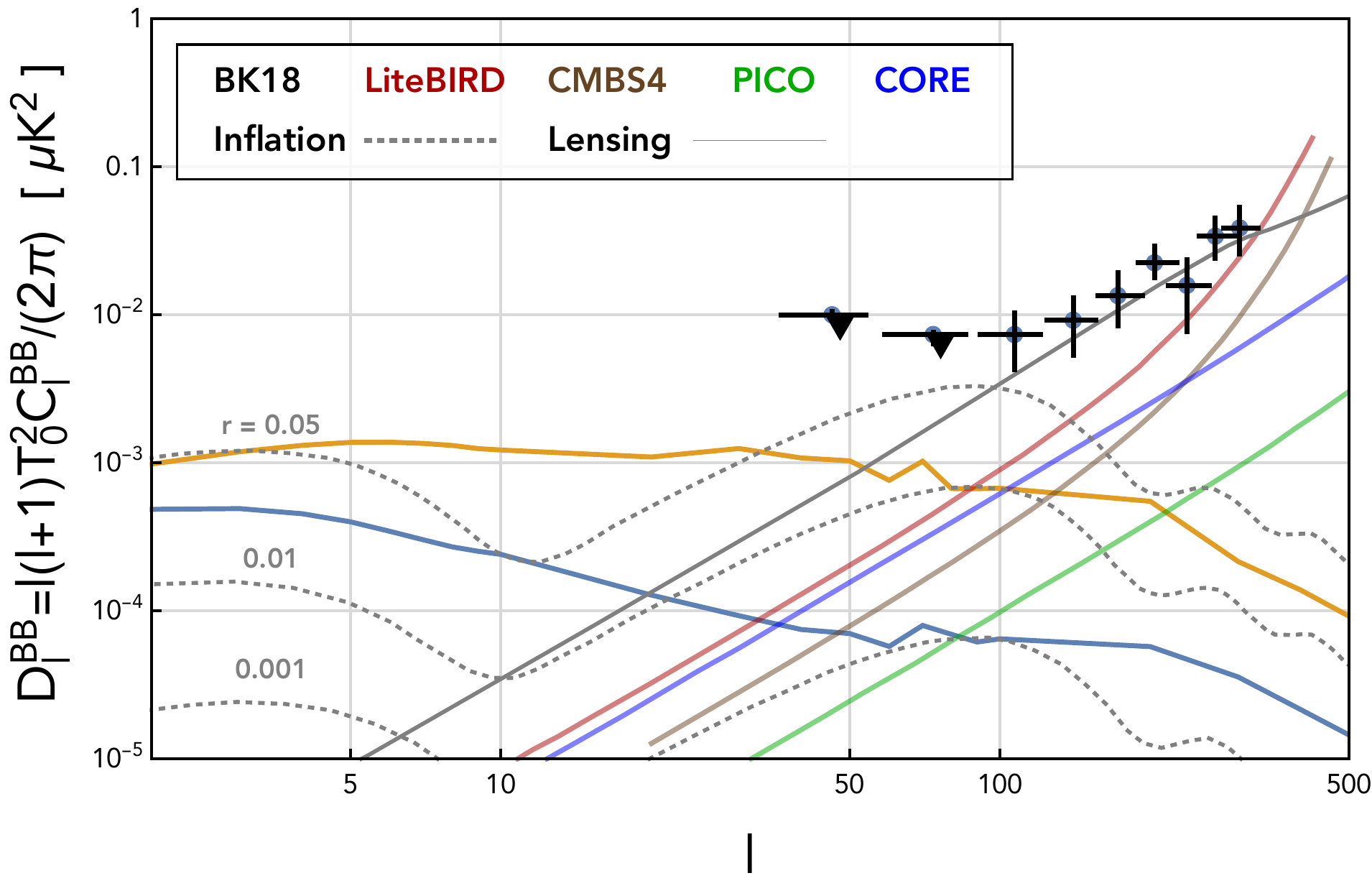} 
\caption{
We plot the CMB power spectrum in our setting, saturating the current Planck 2018 bound ~\cite{Aghanim:2019ame} ({\it left panel}) and show the corresponding B-mode spectra ({\it right panel}) for the two benchmark models (see Table.~\ref{tab:para}). In the left panel we also show the $1\sigma$ error bar of the binned Planck 2018 power spectrum~\cite{Aghanim:2019ame} up to $l=2000$ (when $l>2000$ the uncertainty increases) and in the right panel --  the measurement from BICEP2/Keck Array~\cite{Ade:2015fwj} (digitized from~\cite{Abazajian:2016yjj}) as well as the projected instrumental noise of several future experiments, including LiteBIRD~\cite{Hazumi:2019lys}, CMB-S4~\cite{Abazajian:2016yjj}, PICO~\cite{Hanany:2019lle} and CORE~\cite{Delabrouille:2017rct} (digitized from~\cite{Roy:2020cqn}). Additionally, we plot on the right panel the primordial and lensing B-mode spectra for several different tensor-to-scalar ratio $r$ (dotted gray, taken from~\cite{Hazumi:2019lys}).}
\label{fig:CMB_results}
\end{figure*}

The tensor perturbation $h(\bv{k},\tau)$ is obtained from the linear Einstein equation, which is written in terms of $\bar{h}_{ij}=ah_{ij}$ as
\begin{align}\label{eq:hbar}
\bar{h}^{\prime\prime}_{ij}+\left(k^2-\frac{a^{\prime\prime}}{a}\right)\bar{h}_{ij}=\dfrac{2}{M^2_{Pl}}a\,\Pi_{ij}({\bf k},\tau)\,,
\end{align}
where $\Pi_{ij}({\bf k},\tau)$ is the anisotropic part of the energy momentum tensor $T_{ij}$.
The tensor perturbation then generates the B-mode power spectrum as~\cite{gorbunov2011introduction}
\begin{equation}
\label{eq:ClBB}
\begin{aligned}
C^{BB}_l =& 36\pi \,\mathcal{T}^2_{\rm rei}\int \mc{D}k\mc{D}k^\prime\,\mc{J}^2_{l,B}(k)\\
&\cdot\langle\left\{\int^{\tau_{\rm rei}}_{\tau_{\rm rec}}d\tau\, h^\prime_{ij}(k,\tau)\dfrac{j_2[(\tau_{\rm rei}-\tau)\,k]}{(\tau_{\rm rei}-\tau)^2\,k^2}\right\}^2\rangle\,,
\end{aligned}
\end{equation}
where
\begin{align}\label{eq:j_kl}
\mathcal{J}_{B,l}(k) &= \dfrac{l+2}{2l+1}j_{l-1}(\kappa)-\dfrac{l-1}{2l+1}j_{l+1}(\kappa)\,,
\end{align}
with $\kappa=(\tau_0-\tau_{\rm rei})k$.
We take the narrow width approximation of the visibility function in time as in the $E$-mode calculation. In contrast to the calculation of $TT$, the $B$-mode signal relies on having the last photon scattering at the reionization. This can be seen by the presence of $j_2[(\tau_{\rm rei}-\tau)\,k]$ that comes from expanding the photon propagation within the time interval  $[\tau,\tau_{\rm rei}]$ into spherical harmonics  and then matching the angular mode to the polarization signal.

As can be seen from \eref{eq:hbar}, the spectrum $\langle h^\prime(k,\tau^\prime)h^\prime(k,\tau^{\prime\prime})\rangle$ is related to $\langle\Pi_{ij}({\bf k},\tau^\prime)\Pi_{ij}({\bf k^\prime},\tau^{\prime\prime})\rangle$,  
which again can be expressed in terms of the dark photon mode function $v(k,\tau)$. The B-mode spectrum can therefore be rewritten as
\begin{align}\label{eq:ClBB_final}
\begin{aligned}
C^{BB}_l = &36\pi  \,\mc{T}^2_{\rm rei} \int\mc{D}k\int\mc{D}k_1\Theta(k,k_1,k_2)\\
&\cdot\left(B^2_r(k,k_1,k_2)+B^2_i(k,k_1,k_2)\right)\mathcal{J}^2_{B,l}(k)\,.
\end{aligned}
\end{align}
The function $\Theta$ defined in \eref{eq:Theta_abc} comes from the scalar products of the dark photon polarization, while $B_r$ and $B_i$ are convolutions between $v(k,\tau)$ and the spherical Bessel function $j_2$ (see \eref{eq:Br} and \eref{eq:Bi}). More details of the derivation appear in Appendix~\ref{app:CBB}.

The tensor perturbation also contributes to the $EE$ and $TT$ spectrum. The E-modes have a similar generation mechanism as the B-modes, and we can calculate the $C_{\ell}^{EE}$ by simply replacing  $\mathcal{J}_{B,l}(k)$ in \eref{eq:ClBB} with
\begin{align}
\begin{aligned}
\mathcal{J}_{E,l}(k) = &\dfrac{(l+2)(l+1)}{(2l+1)(2l-1)}j_{l-2}(\kappa)-\dfrac{6(l+2)(l-1)}{(2l+3)(2l-1)}j_l(\kappa)\\
&+\dfrac{l(l-1)}{(2l+3)(2l+1)}j_{l+2}(\kappa)\,,
\end{aligned}
\end{align}
and the changing the pre-factor from $36\pi$ to $9\pi$.
The $TT$ spectrum can be calculated by
\begin{equation}
\label{eq:ClTT_tensor}
\begin{split}
C^{TT}_l &=\dfrac{9\pi}{2}\dfrac{(l+2)!}{(l-2)!}\int \mc{D}k\mc{D}k^\prime\\
&\cdot\langle\left\{\int^{\tau_0}_{\tau_{\rm r}}d\tau\, h^\prime_{ij}(\bv{k},\tau)\dfrac{j_l[(\tau_0-\tau)k]}{(\tau_0-\tau)^2 k^2}\right\}^2\rangle\,.
\end{split}
\end{equation}

As we can see, the power spectra contributed by the tensor and scalar perturbations are proportional $\langle\Pi_{ij}({\bf k},\tau^\prime)\Pi_{ij}({\bf k^\prime},\tau^{\prime\prime})\rangle$, which scales with the mode function of dark photon as $v^4$. The energy density of the dark photon field is $\rho_X=\frac{1}{2a^4}\int\mc{D}k\left(\lvert v^{\prime}(k)\rvert^2+k^2\lvert v(k)\rvert^2-k\right)$, where the last term comes from subtracting the vacuum energy~\cite{Agrawal:2017eqm}. When the mode function grows due to the tachyonic production, the axion initial potential energy  $\rho_\phi=\frac{1}{2}\Lambda^4$ quickly transfers into $\rho_X$ and generates $v\propto \Lambda^2$. 
When fixing the axion mass $m$ and the axion-dark photon coupling $\alpha$, the magnitude of the resulting spectrum is proportional to $\Lambda^8$.

\section{Results: The $TT$ and $BB$ spectra}\label{sec:CMB_results}
In \fref{fig:CMB_results}, we show the $C_{\ell}^{TT}$ and $C_{\ell}^{BB}$ spectra from the two benchmark models defined in Table.~\ref{tab:para}. 
In particular, the value of $\Lambda$ is rescaled 
(keeping $m$ and $\alpha$ fixed) so that the $TT$ spectrum roughly saturates the error bars from the Planck 2018 data \cite{Aghanim:2019ame}, as can be seen in the plot. This shows the rough bounds on $\Lambda$ from the current CMB measurements. On the right of \fref{fig:CMB_results} we plot the corresponding $C_{\ell}^{BB}$ signals for  the two benchmark models saturating the $C_{\ell}^{TT}$ constraints. This gives the upper range of the predicted B-modes within our setting. 

Below we discuss the shape of the calculated spectra. As we can see in \fref{fig:CMB_results} (right), the two axion B-mode curves are roughly parallel to each other. This can be explained by the 
spherical Bessel function
\begin{equation}
x^{-2}j_2(x)\,,\quad x=(\tau_{rei}-\tau)k\,,
\end{equation}
from the angular integral that projects the photon polarization tensor to the B-mode perturbation. The function peaks at the origin and is suppressed by $x^{-3}$ when $x\gg 1$, so the integral is dominated by the $k$-modes that minimize $x$. At the same time, the tachyonic production mainly produces $k$-modes larger than $\tau_{rei}^{-1}$. This results in  $D_{\ell}^{BB}$ getting most of its contribution from perturbations at $\tau\sim\tau_{rei}$. This explains why the difference in the dark photon production at early times between the two benchmarks does not significantly modify the shape of the $\ell$ spectra even though the axions in the two models start rolling at different times (as shown in \fref{fig:V_over_Lambda}). 

This behavior does not hold, however, for the $D_{\ell}^{TT}$ spectra which are sensitive to the starting time of the particle production. 
The $TT$ spectra in Eqs.~(\ref{eq:ClTT_final}) and (\ref{eq:ClTT_tensor}) are not affected by the reionization and the spherical harmonic projection has the form $j_{\ell}[(\tau_0-\tau)k]$, receiving contributions from a wider $\tau$ window for different $\ell$-modes. Our numerical results show that the $TT$ spectra are dominantly contributed by the early period of the dark photon production. This is why they no-longer peak at lower $\ell$-modes as $D_{\ell}^{BB}$, and the peak of the spectrum for the BM1 model, where the particle production starts earlier, is accordingly at higher $\ell$ compared with the peak of the BM2 spectrum.

In the $D_{\ell}^{TT}$ plot, we compare signals from the two benchmark axion models to the Planck 2018 data~\cite{Aghanim:2019ame}, establishing a rough bound on $\Lambda$. We find that $\Lambda_{\rm bound}\approx 15(9)~$meV (see also \tref{tab:para}) for the BM1 (BM2) that saturates the error bar of the Planck data following the same binning as in~\cite{Aghanim:2019ame}. 
We also find a similar sensitivity from the Planck E-mode polarization data, not shown here. 

In the $D_{\ell}^{BB}$ plot, we first note that the BICEP2/Keck measurement~\cite{Ade:2015fwj} does not exclude the benchmark models\footnote{This is not changed by including the lensing effect (gray solid) that shuffles the positions of adiabatic $E$-mode polarization pattern to produce $B$-modes.}. We have accordingly chosen the parameters to satisfy the existing constraints and find that the signal from the late time tachyonic production is well within sensitivities of next-generation CMB B-mode experiments, such as LiteBIRD~\cite{Hazumi:2019lys}, CMB-S4~\cite{Abazajian:2016yjj}, PICO~\cite{Hanany:2019lle} and CORE~\cite{Delabrouille:2017rct}. 
 
 The B-mode signals from axions peak at low-$\ell$, similarly to those from inflationary tensor modes, in both cases due to reionization. In this region, the inflationary model with $r\sim 0.01$ produces B-mode signals that dominate over the gravitational lensing signal (see e.g. Fig. 1 of~\cite{Delabrouille:2017rct}). This suggests that the axion signals can also dominate the lensing background. The scientific goal of LiteBIRD, for example, is to achieve an uncertainty of $\delta r\sim 0.001$ on the range $2\leqslant \ell\leqslant 100$~\cite{Hazumi:2019lys}. It has been shown that even with the contamination from diffuse galactic foreground, LiteBIRD can still be sensitive to $D^{BB}_l\sim 10^{-4}\mu{\rm K}^2$~\cite{Campeti:2020xwn} for $\ell\lesssim 10$. Such sensitivity is close to the BM1 signal, and it is also comparable to the BM2 signal even with a lower $\Lambda\approx 7~$meV,  which is close to the scale of the observed dark energy $\rho_{CC}^{1/4}$. This signal, if observed, might have intriguing implications for the nature of dark energy.

\section{Conclusion}\label{sec:Conclusion}

We have studied the CMB power spectra generated by ALPs via a tachyonic instability and the ensuing production of dark photon quanta close to the cosmic reionization epoch. 
The ALP-dark photon system produces GWs that leave an imprint in the CMB, including its B-mode polarization spectrum. The signal is visible to future CMB polarization detectors while remaining compatible with the bounds from current measurements. Moreover, we find that future experiments can be sensitive to ALP potential energies similar in order of magnitude to the value of CC, which, if discovered, may lead to progress in discerning the nature of dark energy. 
We note that our setting may potentially also generate a signal in measures of cosmic non-Gaussianity that could be visible to future experiments. We leave this analysis for a future study. 

\section{Acknowledgement}
We thank Pedro Schwaller, Ben Stefanek, Chen Sun for useful discussions, and especially Gustavo Marques Tavares for useful comments to the draft. 
MG and SL are supported in part by Israel Science Foundation under Grant No. 1302/19.  MG is also supported in part by the US-Israeli BSF
grant 2018236 and the GIF grant I-2524-303.7. 
YT is supported by the NSF grant PHY-2014165. YT was also supported in part by the National Science Foundation under grant PHY-1914731 and by the
Maryland Center for Fundamental Physics. The authors
also thank the KITP institute (Enervac19 program), and the Munich Institute for
Astro- and Particle Physics (MIAPP) of the DFG Excellence Cluster Origins, where part of this work was conducted, for hospitality.

\bibliography{b-mode}

\onecolumngrid

\appendix
\section{Calculation of the CMB $TT$ spectrum}\label{app:CTT}
Here we give more details about the $C_{\ell}^{TT}$ calculation from the ISW contribution. 
We solve the set of differential equations in Eq.~\eqref{eq:linear_ein} by using the Green's function method and denote the Green's function for $\Phi$ by $G_\Phi(\tau,\tau^\prime)$, which has the boundary condition of $G_\Phi(\tau,\tau)=a(\tau)/(3a^\prime(\tau))$. 
Then for the ISW calculation we should have the conformal time derivative of the gravitational potential $\Phi$ as
\begin{equation}
    \Phi^\prime(\tau)=G_\Phi(\tau,\tau)g(\bv{k},\tau)+\int^\tau_{\tau_{\rm osc}}d\tau^\prime\,\dfrac{dG_\Phi(\tau,\tau^\prime)}{d\tau}g(\bv{k},\tau^\prime)\,,
\end{equation}
where we have defined $g(\bv{k},\tau)\equiv-4\pi G_N\,a^2\delta\rho_e$ for convenience.
With the $d\tau$ integration in \eref{eq:Theta_tilde}, we reorganize the expression by switching the order of the conformal time integrals as
\begin{align}
\tilde{\Theta}_l(\bv{k}) &= 2\int^{\tau_0}_{\tau_{\rm osc}}d\tau\, G_\Phi(\tau,\tau)g(\bv{k},\tau)j_l[k(\tau_0-\tau)]+2\int^{\tau_0}_{\tau_{\rm osc}}d\tau \int^{\tau}_{\tau_{\rm osc}}d\tau^\prime \dfrac{dG_\Phi(\tau,\tau^\prime)}{d\tau}g(\bv{k},\tau^\prime)j_l[k(\tau_0-\tau)]\\
&=2\int^{\tau_0}_{\tau_{\rm osc}}d\tau G_\Phi(\tau,\tau)g(\bv{k},\tau)j_l[k(\tau_0-\tau)]+2\int^{\tau_0}_{\tau_{\rm osc}}d\tau^\prime \int^{\tau_0}_{\tau^\prime}d\tau \dfrac{dG_\Phi(\tau,\tau^\prime)}{d\tau}g(\bv{k},\tau^\prime)j_l[k(\tau_0-\tau)]\\
&=2\int^{\tau_0}_{\tau_{\rm osc}}d\tau^\prime \left\{G_\Phi(\tau^\prime,\tau^\prime)j_l[k(\tau_0-\tau^\prime)] + \int^{\tau_0}_{\tau^\prime}d\tau\, \dfrac{dG_\Phi(\tau,\tau^\prime)}{d\tau}j_l[k(\tau_0-\tau)]\right\}g(\bv{k},\tau^\prime)\\
&= 2\int^{\tau_0}_{\tau_{\rm osc}}d\tau^\prime\, f_{T,l}(k,\tau^\prime)g(\bv{k},\tau^\prime)\,.
\end{align}

We can then convert the correlation function $\langle \tilde{\Theta}_l(\bv{k})\tilde{\Theta}_l(\bv{k^\prime})\rangle$ into $\langle\delta\rho_e(\bv{k},\tau)\delta\rho_e(\bv{k^\prime},\tau^\prime)\rangle$. 
The $\delta\rho_e$ operator can be expressed with the dark photon fields by replacing the $X_i$ and $X_{ij}$ in \eref{eq:delta_rho} with the definitions in \eref{eq:dark_photon}, and its spectrum is obtained as
$\langle0|\delta\left[\mc{O}\right]\delta\left[\mc{O}\right]|0\rangle=\langle0|\mc{O}^2|0\rangle - \langle0|\mc{O}|0\rangle^2$.
After a lengthy but straightforward calculation one arrives at \eref{eq:ClTT_final}. 
The expression of the function $T^r_{\ell}$ and $T^i_{\ell}$ are 
\begin{align}\label{eq:Tr}
T^r_{\ell}(k,k_1,k_2)&=\int^{\tau_0}_{\tau_{\rm osc}}d\tau \dfrac{1}{a^2(\tau)}f_{T,l}(k,\tau){\rm Re}\left\{v_+(\tau,k_1)v_+(\tau,k_2)+\dfrac{v^\prime_+(\tau,k_1)v^\prime_+(\tau,k_2)}{k_1k_2}\right\}\,,\\
\label{eq:Ti}
T^i_{\ell}(k,k_1,k_2)&=\int^{\tau_0}_{\tau_{\rm osc}}d\tau \dfrac{1}{a^2(\tau)}f_{T,l}(k,\tau){\rm Im}\left\{v_+(\tau,k_1)v_+(\tau,k_2)+\dfrac{v^\prime_+(\tau,k_1)v^\prime_+(\tau,k_2)}{k_1k_2}\right\}\,.
\end{align}

\section{Calculation of the CMB B-mode spectrum}\label{app:CBB}

The solution to \eref{eq:hbar} can be written as
\begin{equation}
\bar{h}_{ij}({\bf k},\tau)=\dfrac{2}{M^2_{Pl}}\int^\tau_{\tau_{\rm osc}}d\tau^\prime a(\tau^\prime)G(k,\tau,\tau^\prime)\Pi_{ij}({\bf k},\tau^\prime),
\end{equation}
where $G$ is the Green's function which solves $d^2G/d\tau^2+\left(k^2-a^{\prime\prime}/a\right)G=\delta(\tau-\tau^\prime)$, and satisfies $G(\tau<\tau^\prime)=0$, $G(k,\tau,\tau)=0$ and $G^\prime(k,\tau,\tau)=1$. 
With this expression, the spectrum $\langle h^\prime_{ij}(\bv{k},\tau_1)h^\prime_{ij}(\bv{k},\tau_2) \rangle$ is converted to $\Pi^2({\bf k},\tau^\prime_1,\tau^\prime_2)$, where $\Pi^2({\bf k},\tau^\prime_1,\tau^\prime_2)$ is defined as $\langle\Pi_{ij}({\bf k},\tau)\Pi_{ij}({\bf k^\prime},\tau^\prime)\rangle=(2\pi)^3\Pi^2({\bf k},\tau,\tau^\prime)\delta({\bf k}+{\bf k^\prime})$\,. 
Using the results in Ref.~\cite{Machado:2018nqk}, $\Pi^2({\bf k},\tau,\tau^\prime)$ can be expressed as
\begin{align}
\Pi^2({\bf k},\tau,\tau^\prime)=2\int\mc{D}q\,\Theta_{++}({\bf k}-{\bf q},{\bf k})\mathcal{S}_{++}({\bf q},{\bf k},\tau)\mathcal{S}^*_{++}({\bf q},{\bf k},\tau^\prime)\,,
\end{align}
where the subscript ++ means we include only the positive helicity (which dominates over the negative helicity). 
The function $\Theta$ and $\mathcal{S}$ are also explicitly given in Ref.~\cite{Machado:2018nqk} as
\begin{align}\label{eq:Theta_pp}
\lvert\Theta_{++}(\bv{q},\bv{k})\rvert^2&=\dfrac{1}{16}\left[\left(1+\dfrac{\bv{k}\cdot\bv{q}}{\lvert \bv{k}\rvert\,\lvert \bv{q}\rvert }\right)^2 \left(1+\dfrac{\bv{k}\cdot(\bv{k}-\bv{q})}{\lvert \bv{k}\rvert\,\lvert \bv{k}-\bv{q}\rvert}\right)^2 + \left(1-\dfrac{\bv{k}\cdot\bv{q}}{\lvert \bv{k}\rvert\,\lvert \bv{q}\rvert }\right)^2 \left(1-\dfrac{\bv{k}\cdot(\bv{k}-\bv{q})}{\lvert \bv{k}\rvert\,\lvert \bv{k}-\bv{q}\rvert}\right)^2\right]\,,\\
\label{eq:S_pp}
\mathcal{S}_{++}({\bf q},{\bf k},\tau)&=-\dfrac{1}{a^2(\tau)}\left[\lvert {\bf q}\rvert \lvert {\bf k}-{\bf q}\rvert v_+({\bf q},\tau)v_+({\bf k}-{\bf q},\tau)+v^\prime_+({\bf q},\tau)v^\prime_+({\bf k}-{\bf q},\tau)\right]\,.
\end{align}
These give us all the ingredients for the $C^{BB}_l$ calculation. 
Putting all the explicit expressions back to \eref{eq:ClBB}, and using the same trick as in the calculation of $C^{TT}_l$ to switch the sequence of the two conformal time integrals $\int_{\tau_{\rm osc}}^{\tau_{\rm rei}}d\tau\int_{\tau_{\rm osc}}^{\tau}d\tau'\to\int_{\tau_{\rm osc}}^{\tau_{\rm rei}}d\tau'\int_{\tau'}^{\tau_{\rm rei}}d\tau$ involved in \eref{eq:ClBB}, we arrive at the result \eref{eq:ClBB_final} after a simplification.
The functions involved in the final expression \eref{eq:ClBB_final} are defined as
\begin{align}\label{eq:Theta_abc}
\Theta(a,b,c)&=\dfrac{1}{16}\left[\dfrac{\left((a+b)^2-c^2\right)^2}{4a^2b^2} \dfrac{\left((a+c)^2-b^2\right)}{4a^2c^2} + \dfrac{\left((a-b)^2-c^2\right)^2}{4a^2b^2} \dfrac{\left((a-c)^2-b^2\right)}{4a^2c^2}\right],\\
\label{eq:Br}
B_r(k,k_1,k_2)&=\dfrac{2}{\Mpl^2}\int^{\tau_{\rm rei}}_{\tau_{\rm osc}}d\tau \dfrac{1}{a^2(\tau)}f_B(k,\tau)\,{\rm Re}\left\{v^\prime_+(\tau,k_1)v^\prime_+(\tau,k_2)+k_1 k_2 v_+(\tau,k_1)v_+(\tau,k_2)\right\},\\
\label{eq:Bi}
B_i(k,k_1,k_2)&=\dfrac{2}{\Mpl^2}\int^{\tau_{\rm rei}}_{\tau_{\rm osc}}d\tau \dfrac{1}{a^2(\tau)}f_B(k,\tau)\,{\rm Im}\left\{v^\prime_+(\tau,k_1)v^\prime_+(\tau,k_2)+k_1 k_2 v_+(\tau,k_1)v_+(\tau,k_2)\right\},\\
\label{eq:f_B}
f_B(k,\tau) &= \int^{\tau_{\rm rei}}_{\tau}d\tau^\prime \dfrac{1}{a(\tau^\prime)}\mc{G}(k,\tau^\prime,\tau)\dfrac{j_2[(\tau_{\rm rei}-\tau^\prime)k]}{(\tau_{\rm rei}-\tau^\prime)^2k^2},
\end{align}
where $\mc{G}(k,\tau,\tau^\prime)\equiv dG(k,\tau,\tau^\prime)/d\tau-a^\prime(\tau)/a(\tau) G(k,\tau,\tau^\prime)$.

\end{document}